\documentclass[sigconf,edbt]{acmart-edbt2024}

\def\BibTeX{{\rm B\kern-.05em{\sc i\kern-.025em b}\kern-.08em
    T\kern-.1667em\lower.7ex\hbox{E}\kern-.125emX}}

\usepackage{booktabs} 
\usepackage{graphicx}
\usepackage{subcaption}     

\usepackage{hyperref}
\usepackage{tikz}
\usepackage{pgfplots, pgfplotstable}
\pgfplotsset{compat=1.7}

\usepackage{algorithm,algorithmic, setspace}

\setcopyright{rightsretained}

\acmDOI{}

\acmISBN{978-3-89318-095-0}

\acmConference[EDBT 2024]{27th International Conference on Extending Database Technology (EDBT)}{25th March-28th March, 2024}{Paestum, Italy}

\newcommand{\bwc}{\textsc{BWC} }
\newcommand{\bwcc}{\textsc{BWC}}

\newcommand{\squish}{\textsc{Squish} }
\newcommand{\squishc}{\textsc{Squish}}
\newcommand{\bwcsquish}{\textsc{BWC-Squish} }
\newcommand{\bwcsquishc}{\textsc{BWC-Squish}}

\newcommand{\dr}{\textsc{DR} }
\newcommand{\drc}{\textsc{DR}}
\newcommand{\bwcdr}{\textsc{BWC-DR} }
\newcommand{\bwcdrc}{\textsc{BWC-DR}}

\newcommand{\sttrace}{\textsc{STTrace} }
\newcommand{\sttracec}{\textsc{STTrace}}
\newcommand{\bwcsttrace}{\textsc{BWC-STTrace} }
\newcommand{\bwcsttracec}{\textsc{BWC-STTrace}}
\newcommand{\bwcsttraceopt}{\textsc{BWC-STTrace-Imp} }
\newcommand{\bwcsttraceoptc}{\textsc{BWC-STTrace-Imp}}

\newcommand{\tdtr}{\textsc{TD-TR} }
\newcommand{\tdtrc}{\textsc{TD-TR}}

\begin{document}
\title{New algorithms for the simplification of multiple trajectories under bandwidth constraints}

\author{Gilles Dejaegere}
\orcid{0002-1526-9227}
\affiliation{%
  \institution{Université libre de Bruxelles}
  \streetaddress{Av. Franklin Roosevelt 50}
  \city{Brussels} 
  \state{Belgium} 
  \postcode{1050}
}
\email{gilles.dejaegere@ulb.be}

\author{Mahmoud Sakr}
\affiliation{%
  \institution{Université libre de Bruxelles}
  \streetaddress{Av. Franklin Roosevelt 50}
  \city{Brussels} 
  \state{Belgium} 
  \postcode{1050}
}
\email{mahmoud.sakr@ulb.be}

\renewcommand{\shortauthors}{}

\begin{abstract}
This study introduces time-windowed variations of three established trajectory simplification algorithms. 
These new algorithms are specifically designed to be used in contexts with bandwidth limitations. 
We present the details of these algorithms and highlight the differences compared to their classical counterparts.

To evaluate their performance, we conduct accuracy assessments for varying sizes of time windows, utilizing two different datasets and exploring different compression ratios. The accuracies of the proposed algorithms are compared with those of existing methods. Our findings demonstrate that, for larger time windows, the enhanced version of the bandwidth-constrained \sttrace outperforms other algorithms, with the bandwidth-constrained improved version of \squish also yielding satisfactory results at a lower computational cost. Conversely, for short time windows, only the bandwidth-constrained version of \textsc{Dead Reckoning} remains satisfactory.

\end{abstract}

%
%



\maketitle

\section{Introduction}
During the last decades, the rapid proliferation of mobile devices equipped with tracking capabilities has led to a surge in the production of spatio-temporal data. 
This can be observed across diverse types of geolocation data sources \cite{shekhar2012spatial}. 
Democratization of mobile devices, such as smartphones and wearable technologies, and the spread of Global Positioning System (GPS) equipped vehicles or Automatic Identification System (AIS) equipped vessels are some example of reasons for this data explosion.
While the spatio-temporal data offers many exploitation opportunities (both commercial and research), its increase also causes some new challenges.
One of these challenges is to process this large amount of data.
In 2004, \cite{meratnia2004spatiotemporal} have shown that 100Mb would be necessary to store the localisation of a set of 400 moving objects, with a frequency of 10 Hz (typical frequency of GPS devices).
\textit{Bruxelles Mobilité}\footnote{\href{http://www.bruxellesmobilite.irisnet.be/}{http://www.bruxellesmobilite.irisnet.be/}}, the public administration overseeing mobility-related infrastructure in the Brussels Capital Region, collects positional data specifically for heavy-goods vehicles in Brussels. This information is primarily utilized to calculate toll charges, represents, on average, 19 Gigabytes of data accumulated daily \cite{dillen2020mobi}.

To overcome this difficulty, different compression or simplification algorithms have been proposed \cite{Markis2021, amigo2021review}. 
One of the most well known simplification algorithm is the Douglas Peucker (\textsc{DP}) algorithm \cite{dp}. 
This algorithm was initially aimed at line simplification (without temporal feature). 
Later, \cite{meratnia2004spatiotemporal} introduced some variations of the \textsc{DP} algorithm (including the \textsc{Top Down Time Ratio} algorithm (\tdtrc)), taking into account the temporal feature of the locations.
Since then, multiple algorithms such as \squish (and its variations) \cite{squish, squishe}, \sttrace \cite{potamias2006sampling} or \textsc{Dead Reckoning} (\drc) \cite{trajcevski2006line} have been proposed.
The main contribution of this work is to extend the \squishc, \sttrace and \dr algorithms so that they could be used in contexts where bandwidth limitations apply.
The rest of this paper is divided as follows. 
First, Section~\ref{sec:motivation} will provide a definition of compression under bandwidth constraints as well as a motivation to this problem. 
Section~\ref{sec:classic_algorithms} introduces three existing trajectory simplification algorithms while 
variants of these algorithms for the bandwidth constrained contexts are described in Section~\ref{sec:BWC}. 
Then, in Section~\ref{sec:results}, the performances of these algorithms will be analysed and compared to the existing algorithms using two different datasets. 
It will also be shown that the classical algorithms are not suited for bandwidth constrained contexts.
Finally, Section~\ref{sec:ccl} concludes this work and presents some further research avenues.

\section{Compression under bandwidth constraints motivation} \label{sec:motivation}

Existing techniques for simplification of trajectories have already largely been studied. 
These techniques are generally aimed at simplifying the trajectories in order to facilitate their exploitation by machine learning techniques. 
This is usually performed by trying to minimize the number of points (position of an object at a given timestamp) kept without deteriorating the trajectory significantly.
In this work, a different approach will be used. 
Instead of trying to minimize the number of points kept, the algorithms introduced in this work will consider some bandwidth constraints.
These constraints are defined as follows. 
For each period of time, bandwidth constrained algorithm must respect a predefined limit on the quantity of points that can be kept.
Therefore algorithms presented in this work are aimed at minimizing the deterioration of the trajectories during compression without exceeding this limit on the quantity of points kept, and this, for all time periods.
The size of these periods as well as the number of points that can be kept are parameters of the compression algorithms.
While bandwidth limitations are mentioned for different contexts (vessels tracking \cite{mcgillivary2009enhancing}, animal tracking \cite{juang2002energy}),
the problem of simplifying trajectories under bandwidth limitation has, to the best of the authors knowledge, not yet attracted the attention of the research community.
Some existing algorithms (such as the already mentioned \squish and \sttracec) provide some solutions to compress trajectories under memory limitations (the final number of points is a parameter of the methods) but these are not adapted for bandwidth constrained contexts.

The main use case motivating compression of trajectories under bandwidth constraints concerns the extension of AIS signal coverage for maritime monitoring and is detailed in Section \ref{sec:ais_coverage}.
Further potential use cases are detailed in Section~\ref{sec:other_use_cases}.

\subsection{Extension of AIS signal coverage} \label{sec:ais_coverage}

Since 2004, all cargo vessels over 500 GT and all passenger vessels are required to be equipped
with AIS transceivers. 
These transceivers allow automatic exchange of information in between ships and between ships and coastal stations by broadcasting positional messages using the Self-Organizing Time Division Multiple Access (SOTDMA) protocol.
The International Telecommunication Union (ITU) recommendation defines 2 default communication frequencies:
AIS 1 (161.975 MHz) and AIS 2 (162.025 MHz) \cite{series2014technical}.

While AIS data has initially been developed for collision avoidance, since then, it has vastly been used by maritime authorities to monitor vessels' behavior and identify illegal activities.
The frequencies and the use of SOTDMA protocol imposed by the ITU however limit both the range of communication and the bandwidth available.
One vastly used solution to increase the range for which vessels could be monitored from coastal stations is satellite AIS.
Satellite AIS involves the use of satellites to receive and relay AIS signals from the ships to the stations.
Another possible solution which does not require the use of satellites is mentioned in \cite{mobispaces}.
It consists in allowing ships to repeat some of the broadcasted signals that they receive from each other, acting as an "AIS-repeater".
Such a solution however would come at the cost of an increase in the size of data transmission, which, if applied naively, might exceed the available bandwidth.
For this reason, compression techniques adapted to bandwidth constraints should be developed.

\subsection{Objects tracking over the Internet of Things} \label{sec:other_use_cases}

Another family of use cases where trajectory compression under bandwidth constraints might be beneficial could be the tracking of objects over the Internet of Things (IoT).
By design, many IoT devices have limited capabilities (battery, bandwidth, ...). 
Object tracking devices with such limitations could need to compress the trajectories before communicating them to other devices.
For such devices, compression is not aimed at but a technical necessity.
In this situation bandwidth constrained compression algorithms would offer the necessary compression while minimizing the deterioration of the trajectories.
Many situations could be considered. 
Some examples are given as follows:
\begin{description}
        \item[Animal tracking:] Animal tracking is more and more used by private pet owners, live stock owners and by scientists. 
                For the latter, compressing animals' trajectories under bandwidth constraints might be necessary to study animals' behaviors in remote locations where communication capabilities are inherently constrained.
        \item[Autonomous fleets:] with the recent development of smart cities and autonomous vehicles, the amount of positional information generated is always increasing.
                Combined with the additional information exchange required in such smart-cities, the monitoring of the trajectories of fleets of autonomous vehicles might therefore benefit from bandwidth constrained compression.
\end{description}


%
%
%
%
\section{Existing algorithms} \label{sec:classic_algorithms}

In this section, 3 existing algorithms which can be adapted in bandwidth constraint scenarios will be introduced.
For all these algorithms, we will consider $n$ entities (or targets) for which the position on earth is tracked over time.
For each entity $l$, its actual continuous movement over time will be called its real trajectory and denoted by $\mathcal{T}_l$.
In practice, this continuous trajectory will be measured at discrete timestamps leading to the generation of the trajectory of $l$, denoted $t_l$ as a time ordered sequence of measurements of $l$'s position.
 
The main purpose of the algorithms will be to compress (or simplify) the $n$ trajectories into $n$ samples (denoted $s_l$ with $l\in \{1,...n\})$.
In this work, we will only consider compression techniques such that the sample $s_l$ obtained by compressing $t_l$ is composed of a subset of the points of $t_l$ (still ordered by their timestamps).

In addition for being important algorithms in the trajectory compression literature, these three algorithms have been chosen for the following reason.
Both \squish and \sttrace are designed to compress trajectories to a predetermined target size which inherently makes them suitable candidates to be adapted in a bandwidth constrained context.
\drc, on the other hand, is inherently designed to be applied in real time and will be modified in order to be able to respect bandwidth limitations.

Hereunder, the three classical algorithms will be introduced.
For simplicity purposes, \squish and \sttrace will be illustrated with a priority queue. 
It should be noted however that this is done to simplify the algorithms description and that the priority queue is not an inherent characteristic of the methods. 
Indeed, both methods could be implemented more efficiently without it.

\subsection{\squish}

The \squish algorithm has initially been presented in \cite{squish}. 
Since then, several improvements have been proposed, such as the \textsc{Squish-E} method presented in \cite{squishe}.
It works by compressing each trajectory individually.
It will therefore receive as input a single trajectory.
Each point $p$ in this trajectory will be a tuple composed of $(p.x, p.y, p.ts)$ with 
$p.x$ and $p.y$ being its coordinates and $p.ts$ being the timestamp associated to $p$.
Furthermore, the algorithm will associate to each point a dynamic priority $p.priority$ which will depend on the current state of the sample $s$ representing the trajectory.
The main steps of the algorithm are described in Algorithm~\ref{algo:squish}.

\begin{algorithm}[ht]
\begin{algorithmic}[1]
        \REQUIRE trajectory $t$, sample size $M_t$
    \STATE $s$ = empty list of points
    \STATE $\mathcal{Q}$ = empty priority queue 
    \STATE
    \FOR{$p$ in $t$}
        \STATE p.priority = $\infty$
        \STATE $s$.append($p$)
        \STATE compute\_priority($s[-2]$, $s$) \COMMENT{s[-2] is the previous point} \label{l:compute_priority_squish}
        \STATE $\mathcal{Q}$.add($p$) 
        \IF{$\mathcal{Q}$.size() > $M_t$}
            \STATE drop\_point\_update\_priorities($\mathcal{Q}$, s) \label{l:drop_point_squish}
        \ENDIF
    \ENDFOR
    \RETURN $s$ \COMMENT{or $\mathcal{Q}$}
\end{algorithmic}
\caption{Pseudocode of the \squish algorithm}
\label{algo:squish}
\end{algorithm}

The priority of a point in the sample (line \ref{l:compute_priority_squish}) is computed as the \textit{Synchronized Euclidian Distance} (SED) error introduced in the sample by removing this point. 
The SED of a point $x$ with respect to points $a$ and $b$ such that:
\begin{equation}
a.ts \le x.ts \le b.ts
        \label{eq:point_order}
\end{equation}
represents the distance between the point $x$ and its projection $x^\prime$ which is the position the entity would have at time $x.ts$ if it was moving at constant speed between $a$ and $b$.
Therefore, $SED(a, x, b)$ be computed as follows:
\begin{equation}
        SED(a, x, b) = dist(x, pos(a, b, x.ts))
        \label{eq:SED}
\end{equation}
with the distance between two points being computed as their euclidian distance:
\begin{equation}
        dist(a,b) = \sqrt{(a.x - b.x)^2 + (a.y - b.y)^2}
        \label{eq:distance}
\end{equation}
and with the position at a specific time $time\in [a.ts, b.ts]$ (according to a segment between the two other points $a$ and $b$) being defined by:
\begin{align}
        pos(a, b, time).x & = a.x + \frac{(b.x - a.x)}{b.ts-a.ts}\times (time - a.ts) \label{eq:position_x} \\ 
        pos(a, b, time).y & = a.y + \frac{(b.y - a.y)}{b.ts-a.ts}\times (time - a.ts) \label{eq:position_y}
\end{align}
The priority of a point at the position $l$ in a sample $s$ of size $k$ is computed as follows:
\begin{align}
        \begin{split}
        compute\_priority(s[l], s) = & \quad SED(s[l-1], s[l], s[l+1]) \\
                                     & \qquad \qquad \qquad \forall l=1,..., k-1 
        \end{split}
        \label{eq:priority}
\end{align}
With the priorities of $s[0] = s[k] = \infty$ as the first and the last point of the sample will always be kept.

When a new point is added to the sample, the size of the priority queue might exceed the maximum allowed buffer size.
In this case, the point with the lowest priority should be dropped (both from the sample and from priority queue) (see line \ref{l:drop_point_squish}).
Once a point is dropped, the priority of the "neighbors" of this point should be updated.
In order not to recompute the priority of the points, \squish works by increasing the priority of the neighboring points by the priority of the point dropped.
By denoting $s$ the sample before the dropping of the point $s[l]$ and $s^\prime$ the sample after the removal, the priorities of the points which were neighboring $s[l]$ will be computed as follows:
\begin{align}
        s^\prime[l-1].priority &= s[l-1].priority + s[l].priority \\
        s^\prime[l].priority &= s[l + 1].priority + s[l].priority
        \label{eq:squish_priority_update}
\end{align}
It should be noted that the point following $s[l]$ has the index $l+1$ in $s$ while it has the index $l$ in $s^\prime$ due to the removal of $s[l]$.
Once the priority of these two points is recomputed, their positions in the priority queue are adapted as well.

It is important to keep in mind that for \squish as well as for all all other algorithms presented in this work, when a point is dropped, it is dropped both from the priority queue and from the sample it belongs to.

\subsection{\sttrace}

The \sttrace algorithm was initially presented in \cite{potamias2006sampling}.
Its pseudocode is presented in Algorithm \ref{algo:STTrace}.

\begin{algorithm}[ht]
\begin{algorithmic}[1]
    \REQUIRE $Stream$ $\mathcal{S}\mathcal{T},$ maximal buffer size $Mn$ 
    \STATE $S$ = matrix of $l$ empty lists 
    \STATE $\mathcal{Q}$ = empty priority queue 
    \STATE
    \FOR{$p$ in $\mathcal{S}\mathcal{T}$}   \label{l:input}
        \STATE s = S[p.id]
        \IF{interesting(p, s, $\mathcal{Q}$)} \label{l:interesting}
                \STATE p.priority = $\infty$
                \STATE $s$.append($p$)
                \STATE compute\_priority($s$, $s[-2]$) \label{l:compute_priority_sttrace}
                \STATE $\mathcal{Q}$.add($p$) 
                \IF{$\mathcal{Q}$.size() > $M_t$}
                    \STATE drop\_point\_recompute\_priorities($\mathcal{Q}$, S) \label{l:drop_point_sttrace}
                \ENDIF
        \ENDIF
    \ENDFOR
    \RETURN $S$ \COMMENT{or $\mathcal{Q}$}
\end{algorithmic}
\caption{Pseudocode of the \sttrace algorithm}
\label{algo:STTrace}
\end{algorithm}

It is very similar to \squish except the three following differences:
\begin{description}
    \item[line \ref{l:input}:] It compresses the different trajectories simultaneously (the $n$ trajectories are contained into a single stream of points $\mathcal{S}\mathcal{T}$). 
    Each point $p$ of the stream will be a tuple composed of $(p.id, p.x, p.y, p.ts)$ with $p.id$ being the index of the trajectory $t_{p.id}$ it belongs to, 
    $p.x$ and $p.y$ being its coordinates and $p.ts$ being the timestamp associated to $p$.

            Furthermore, it operates in an unbalanced way, i.e. after simplification, samples representing more complicated trajectories will be composed of more points than samples representing more simple ones. 
            This result is obtained by maintaining a single priority queue for all the points of the different trajectories.
    \item[line \ref{l:drop_point_sttrace}:] When one point $x$ is dropped from the priority queue and from the concerned sample $S[x.id]$ (note that the sample $S[x.id]$ is generally not the same sample as the sample in which the last point was added), the priorities of the neighboring points of $x$ in the sample $S[x.id]$ will not be updated using an heuristic approach such as in \squishc. 
     Instead, when removing a point $s[l]$ from the sample $s$, both the priority of $s[l-1]$ will be recomputed 
     as $SED(s[l-2], s[l-1], s[l+1])$ and the priority of $s[l+1]$ will be recomputed 
     as $SED(s[l-1], s[l+1], s[l+2])$.

    \item[line \ref{l:interesting}:] Before adding the next point $p$ in a sample $s=S[p.id]$, it will first check whether this point seems promising.
            This is performed by computing what the priority of the last point in $s$ would be if $p$ was added to $s$ :\linebreak $SED(s[-2], s[-1], p)$.
            If this potential priority is lower than the lowest priority in the priority queue, then point $p$ is not added to the sample.
\end{description}

\subsection{\dr} \label{sec:DR}

The \dr algorithm has been initially presented in \cite{trajcevski2006line}. 
It has the particularity of being inherently designed for real-time applications.
The main idea is that when a point $p$ is considered to be added to the sample $s$, the deviation between $p$ and the expected position according to the last points of $s$ at the time $p.ts$ will be computed. 
If this deviation is larger than a defined threshold, then $p$ is added to the sample $s$.
The pseudocode for the \dr algorithm is provided in Algorithm~\ref{algo:dr}

\begin{algorithm}[ht]
\begin{algorithmic}[1]
    \REQUIRE $Stream$ $\mathcal{S}\mathcal{T},$ deviation threshold $\epsilon$ 
    \STATE $S$ = matrix of $l$ empty lists 
    \FOR{$p$ in $\mathcal{S}\mathcal{T}$}   
        \STATE s = S[p.id]
        \STATE $p^\prime$ = estimate\_position(s, p.ts) \label{l:expected_pos}
        \IF{dist($p^\prime, p$) > $\epsilon$} 
                \STATE $s$.append($p$)
        \ENDIF
    \ENDFOR
    \RETURN $S$ 
\end{algorithmic}
\caption{Pseudocode of the \dr algorithm}
\label{algo:dr}
\end{algorithm}

The estimated position (line \ref{l:expected_pos}) can be computed in two different ways according to the information contained in the stream of points.
If each point $p$ of the stream is composed of \newline $(p.id, p.x, p.y, p.ts)$ (which is also the information required by the \squish and \sttrace algorithms), then the expected position will be computed as if the object was travelling with constant direction and speed from $s[-1]$ (with the direction and the speed being computed according to the straight line between $s[-2]$ and $s[-1]$):
\begin{align}
        p^\prime.x & = s[-1].x + \frac{(s[-1].x - s[-2].x)}{s[-1].ts-s[-2].ts}\times (p.ts - s[-1].ts) \\
        p^\prime.y & = s[-1].y + \frac{(s[-1].y - s[-2].y)}{s[-1].ts-s[-2].ts}\times (p.ts - s[-1].ts) 
        \label{eq:expected_position}
\end{align}

In some cases (such as in the AIS data), each point $p$ in the stream contains some information with respect to its speed and direction of the moving object.
Each point p is then composed of $(p.id, p.x, p.y, p.ts, p.sog, p.cog)$ with $p.sog$ and $p.cog$ representing respectively the speed over ground and course over ground of the entity.
Then this additional information can be used to compute the estimated position $p^\prime$ of $p$:
\begin{align}
        p^\prime.x & = s[-1].x + cos(s[-1].cog)\times s[-1].sog\times (p.ts - s[-1].ts) \\
        p^\prime.y & = s[-1].y + sin(s[-1].cog)\times s[-1].sog\times (p.ts - s[-1].ts) 
        \label{eq:expected_position_sog}
\end{align}

\section{\bwc variants} \label{sec:BWC}

While the previous section consisted in an introduction of different existing compression techniques, this section consists in the introduction of \textit{BandWidth-Constrained} (\bwcc) variants of the existing algorithms.
Four variants will be analysed in this work: \bwcsttracec, \bwcsttraceoptc, \bwcsquishc, \bwcdrc.
All of them share the main idea of extending their respective existing algorithm in a time windowed manner.
However some slight adaptations have to be performed for the \bwcsquish and \bwcdr algorithms.
Furthermore, the time windowed constraint also gives us the opportunity of proposing ``improvement" of the \bwcsttrace algorithm (which is denoted \bwcsttraceoptc).
The modifications necessary for these three algorithms will be developed hereunder.

For simplicity purposes, the bandwidth will be considered as a constant parameter in all the algorithms.
This means that for each time window, the same number of points will be kept.
However, in practice, nothing prevents the algorithms of being used with an array of bandwidths for each different time window or in a more dynamic way by adapting the bandwidth according to the real time congestion of the network.

\subsection{\bwcsquish and \bwcsttrace} \label{sec:BWC_Squish}
The \bwcsttrace method is simply the modification of the \sttrace method applied on every time window, with the particularity that points kept in the sample of previous time windows can be used to compute the priority of points in the current time window.
The priority of points in \bwcsttrace is identically computed as in the original \sttrace method.
The bandwidth constrains are respected by flushing and re-initializing the priority queue after each time window. 
A similar approach is used for the \bwcsquish algorithm.
One of the characteristics of the \squish method, is that the numbers of points kept in the simplification of the trajectories have to be determined beforehand. 
However, the repartition of the number of points that should be kept for each trajectory individually in each time window is not straight forward. 
For this reason, the \bwcsquish algorithm is an ``\sttrace inspired" modification of the \squish algorithm as instead of compressing the trajectories individually, a single priority queue of limited size is shared for all trajectories.
Such as for \bwcsttrace, the priority of points in \bwcsquish is identically computed as in the original \squish method.
The pseudocode for the algorithms fo \bwcsttrace and \bwcsquish are identical and are shown in Algorithm~\ref{algo:bwc-sttrace-opt}.
While the pseudocodes are identical, it is important to remember that both methods still compute the priorities differently.

\begin{algorithm}[ht]
\begin{algorithmic}[1]
    \REQUIRE $Stream$ $\mathcal{S}\mathcal{T},$ window limit $bw$, window duration $\delta$, start time $start$, \underline{precision $\epsilon$}
    \STATE $S$ = matrix of $l$ empty lists 
    \STATE \underline{$T$ = matrix of $l$ empty lists }
    \STATE $\mathcal{Q}$ = empty priority queue 
    \STATE $window\_end$ = $start + \delta$
    \FOR{$p$ in $\mathcal{S}\mathcal{T}$}   
        \IF{$p.ts > window\_end$} 
                \STATE flush($\mathcal{Q}$) 
                \STATE $window\_end$ = $window\_end + \delta$
        \ENDIF
        \STATE s, \underline{t} = S[p.id], \underline{T[p.id]}
        \STATE p.priority = $\infty$
        \STATE $s$.append($p$)
        \STATE \underline{$t$.append($p$)}
        \STATE compute\_priority\_imp($s[-2]$, $s$, \underline{$t$, $\epsilon$}) 
        \STATE $\mathcal{Q}$.add($p$) 
        \IF{$\mathcal{Q}$.size() $> bw$}
                \STATE drop\_point\_recompute\_priorities($\mathcal{Q}$, S, \underline{T, $\epsilon$}) 
        \ENDIF
    \ENDFOR
    \RETURN 
\end{algorithmic}
\caption{Pseudocode of the \bwcsquish, \bwcsttrace and \bwcsttraceopt algorithms. Underlined parts are the addition required for \bwcsttraceopt.}
\label{algo:bwc-sttrace-opt}
\end{algorithm}


\subsection{\bwcsttraceopt} \label{sec:BWC_STTrace_Imp}

The main motivation behind this improvement is that in \sttracec, the priority of a point is computed using the sample it belongs to.
Therefore, this priority is computed independently of the previously removed points. 
While the removal of a single point with a small priority will lead to a slight deviation in the sample, significant deviations can result of successively removing such points.
The pseudocode of \bwcsttraceopt is detailed in Algorithm~\ref{algo:bwc-sttrace-opt}. 

The priority of a point in a sample is therefore computed as follows.
Instead of computing the SED error introduced in the sample when removing the concerned point, 
\bwcsttraceopt computes the difference between the SED error of the sample with respect to the initial trajectory with and without the considered point. 

This error will be computed according to the distance between the synchronized position in the trajectory and the position in corresponding sample at regular time intervals (denoted $\epsilon$).
To compute these positions (in a trajectory or in sample denoted $x$) at a specific time $t$, the "neighboring" points should be identified.
These neighbor points will be denoted $x^{-}_t$ (the first point in $x$ before time $t$) and $x^{+}_t$ (the first point in $x$ after time $t$):
\begin{align}
        \begin{split}
        x^{-}_t =\quad &p \in x \quad s.t. \\
                  & p.ts \le t \\
                  & \wedge \not \exists q \in x \quad s.t. \quad  p.ts < q.ts \le t
        \end{split} \\
        \begin{split}
        x^{+}_t =\quad &p \in x \quad s.t. \\
                  & t \le p.ts  \\
                  & \wedge \not \exists q \in x \quad s.t. \quad     t \le q.ts < p.ts 
        \end{split}
        \label{eq:neighboors}
\end{align}

By using equations \ref{eq:position_x}, \ref{eq:position_y} and \ref{eq:neighboors}, we will define a function $x(t)$ providing the position of the entity at time $t$ according to the sample or trajectory $x$:
\begin{equation}
        x(t) = pos(x^{-}_t, x^{+}_t, t)  
        \label{eq:position_in_sample}
\end{equation}
Then, the set of all the timestamps where the errors will be computed will be denoted $W(s[l], s)$.
Indeed, the priority of a points $s[l]$ will be the sum of all the errors for all timestamps between $s[l-1].ts$ and $s[l+1].ts$ with the step $\epsilon$.
$W(s[l], s, \delta)$ will therefore be denoted:
\begin{multline}
        W(s[l], s, \delta) = \{s[l-1].ts + k\epsilon \mid  \\
        k\in \mathbb{N}^+ \wedge s[l-1].ts + k\epsilon < s[l+1].ts\} 
        \label{eq:timespawn}
\end{multline}
Finally, the sample that would be obtained by removing the node $s[l]$ from $s$ will be denoted:
\begin{equation}
        s^{-l} = s \setminus s[l]
        \label{eq:new_sample}
\end{equation}

Using these notations, the priority in the sample $s$ with respect to an initial trajectory $traj$ of a point $s[l]$ can then be computed as:
\begin{multline}
                compute\_priority\_imp(s[l], s, traj, \epsilon) = \hfill  \\
        \sum\limits_{\substack{t \in W(s[l], s, \epsilon)}} \Big(dist\big(traj(t), s(t))\big)
                                                                       - dist\big(traj(t), s^{-l}(t))\big)\Big)
        \label{eq:priority_opt_reg}
\end{multline}

Once more, such as in \sttracec, when dropping a point from a sample, the priority of the previous and following points in the sample will need to be recomputed.

While \bwcsttraceopt will produce more accurate results, it is at the cost of a more computationally expensive computation of the priorities.
The computation of the priority of the point $s[l]$ in \sttrace or \bwcsttrace requires the computation of one distance as well as one position (from two existing points and one timestamp).
On the other hand the computation of the priority of $s[l]$ in \bwcsttraceopt requires the computation of at most $\frac{2\times\delta}{\epsilon}\times 2$ distances as well as $\frac{2\times\delta}{\epsilon}\times 3$ positions.
Indeed, since $s[l-1]$ might belong to the previous time window, the duration between $s[l-1]$ and $s[l+1]$ is at most $2\times \delta$ which leads the set $W(s[l], s, \epsilon)$ to be at most of size $\frac{2\times\delta}{\epsilon}$. 
For every timestamp in this set, $3$ positions (according to the real trajectory, the initial sample and the simplified sample) as well as $2$ distances must be computed.


\subsection{\bwcdr} \label{sec:BWC_DR}

The \dr algorithm has been modified in order to fulfill bandwidth constrains.
This is performed, such as for \squish and \sttrace by the introduction of time windows and a priority queue.
Instead of using the distance between the position of the processed point with its expected position as a binary criterion to decide whether to add this point to the corresponding sample or not, this distance will be used as the priority of the point.
Therefore, only the points which are the furthest of their expected position will be kept in each time window.

The pseudocode for the \bwcdr algorithm is detailed in Algorithm~\ref{algo:bwc-dr}.

\begin{algorithm}[ht]
\begin{algorithmic}[1]
    \REQUIRE $Stream$ $\mathcal{S}\mathcal{T},$ window limit $bw$, window duration $\delta$, start time $start$ 
    \STATE $S$ = matrix of $l$ empty lists 
    \STATE $\mathcal{Q}$ = empty priority queue 
    \STATE $window\_end$ = $start + \delta$
    \STATE
    \FOR{$p$ in $\mathcal{S}\mathcal{T}$}   
        \IF{$p.ts > window\_end$} 
                \STATE flush($\mathcal{Q}$) 
                \STATE $window\_end$ = $window\_end + \delta$
        \ENDIF
        \STATE $p^\prime$ = estimate\_position(s, p.ts) 
        \STATE p.priority = dist($p^\prime, p$) 
        \STATE $s$.append($p$)
        \STATE $\mathcal{Q}$.add($p$) 
        \IF{$\mathcal{Q}$.size() > $bw$}
                \STATE drop\_point\_recompute\_priorities($\mathcal{Q}$, S) 
        \ENDIF
    \ENDFOR
    \RETURN 
\end{algorithmic}
\caption{Pseudocode of the \bwcdr algorithm.}
\label{algo:bwc-dr}
\end{algorithm}

Similarly as with \squish and \sttrace (bandwidth constrained versions or not), when one point $s[l]$ is dropped from the priority queue, it is also removed from the corresponding sample.
Therefore, the priorities of some points of $s$ must be recomputed.
With \bwcdrc, its is not the priorities of the two neighbors ($s[l-1]$ and $s[l+1]$) which must be recomputed, but the priorities of the one or two next nodes ($s[l+1]$ and $s[l+2]$). 

\section{Empirical results} \label{sec:results}

In this section, the performance of the introduced \bwc algorithms as well as their classical equivalents and the classical \tdtr algorithm will be compared.
The comparison will be performed on two datasets of different spatial and temporal ranges.

\subsection{Datasets}

\subsubsection{AIS}
\begin{figure}[ht]
        \begin{center}
                \includegraphics[width=0.45\textwidth]{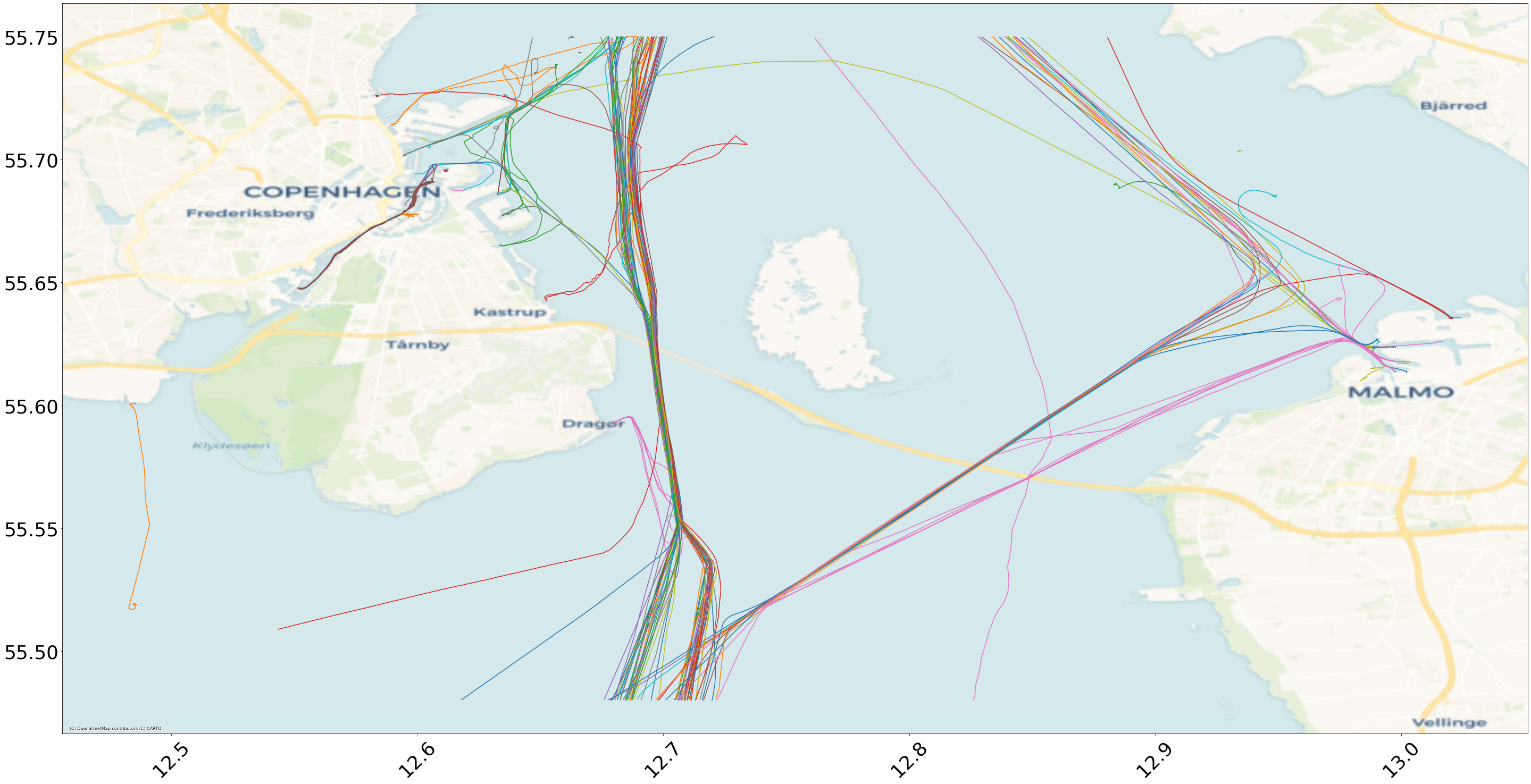}
        \end{center}
        \caption{AIS trips around Copenhagen and Malmo.}\label{fig:copenhagen}
\end{figure}

The first dataset consists of 24h of AIS data in the region between the cities of Copenhagen and Malmo on first January 2021 \cite{AISDenmark}.
It is composed of 103 trips totalling 96819 points. 
The trips can be seen in Figure~\ref{fig:copenhagen}.

\subsubsection{Birds}

\begin{figure}[ht]
        \begin{center}
                \includegraphics[width=0.35\textwidth]{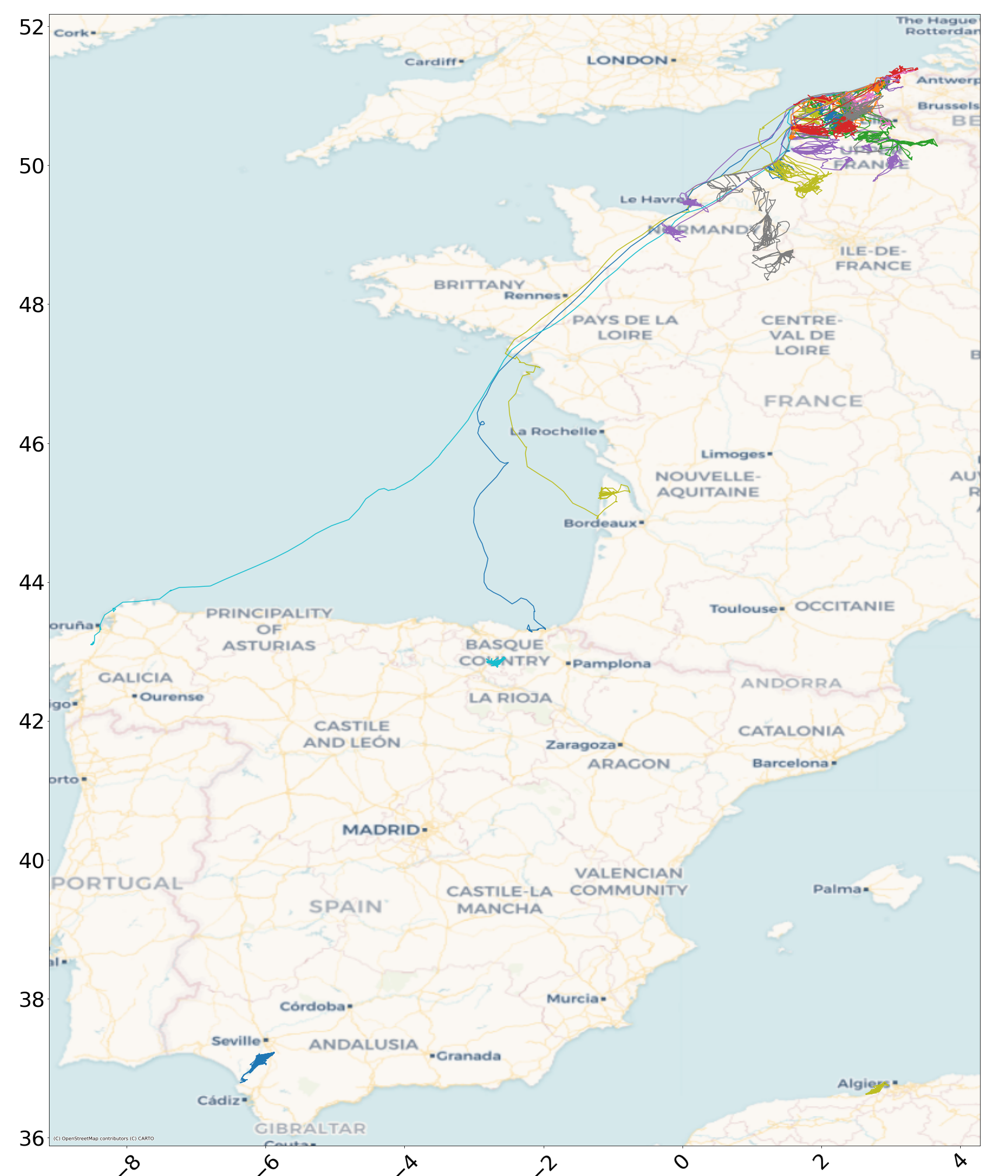}
        \end{center}
        \caption{Birds trips.}\label{fig:birds}
\end{figure}

The second dataset consists of three months of GPS of black-backed gulls between the 9th of July and the 9th of October 2021 \cite{birds}.
It is composed of 45 trips totalling 165244 points. 
While most of these trips originate from Belgium and North of France, some are spreading as far as the north of Spain. 
Few other trips are also entirely taking place in Spain and one in Algeria.
These trips can be seen in Figure~\ref{fig:birds}.

\subsection{Algorithm accuracy}

In this section, the accuracy of the different algorithms is computed. 
In order to asses this accuracy, every algorithm is applied to simplify all initial trajectories. 
Once these simplifications are computed, the distance between synchronized projection on the initial and simplified trajectories are computed at a regular time interval.

It is important to note that this accuracy evaluation is not aimed at stating that some compression algorithms are better than others. 
Indeed, when selecting a compression algorithm, different factors have to be taken into account.
While the accuracy is generally an important factor, other factors such as time and space complexity should be taken into account.
Furthermore the \bwc algorithms are designed to be able to be used in situations with additional bandwidth constraints.
It is not surprising that the fulfillment of these additional constraints may lead to a deterioration of the algorithms accuracy.

While the different algorithms require different parameters, these were determined in order to produce a similar total number of points in the simplified trajectories produced by the different algorithms.
For each dataset, the algorithms will be assessed with parameters such that both around $10\%$ and around $30\%$ of the original points are kept in the simplified trajectories.

The exact value of the parameters for the classical algorithms are listed hereunder. 
\begin{description}
        \item[\squish] \squish requires the maximal number or points kept for each individual trajectory. This maximal number of point as been set to $10\%$ and $30\%$ of the initial points of each trajectory.
        \item[\sttrace] \sttrace requires the maximal number or points kept for for all trajectories. This maximal number of points has been set to $10\%$ and $30\%$ of all initial points.
        \item[\dr] \dr requires a distance threshold. 
                This threshold has been set to 425 and 115 meters for the ais dataset and has been set to 2500 and 950 meters for the birds dataset.
        \item[\tdtr] The \tdtr time algorithm requires a tolerance threshold. This threshold has been set to 0.15 and 0.051 in the AIS dataset as well as 16.7 and 1.5 for the Birds dataset.
\end{description}

For each of the classical algorithms, its accuracy (average distance in meters between the simplified and original trajectories) can be seen in Table~\ref{tab:results_classical}.
\begin{table}[ht]
\begin{tabular}{l r r r r}  
\toprule
                     & \multicolumn{2}{c}{AIS}            & \multicolumn{2}{c}{Birds} \\ 
\cmidrule(r){2-3}\cmidrule(r){4-5}
                     & 10\%  &  30\% &  10\% & 30\% \\
\midrule                                 
\squish              &  20.87  &   4.83       &  585.34  &  44.95        \\
\sttrace             &  58.66  &   9.78       & 1823.10  & 431.65       \\
\dr                  &  42.68  &  13.12       &  697.14  &  46.48        \\
\tdtr                &   2.95  &   1.08       &  274.78  &  26.87        \\
\bottomrule
\end{tabular}
\caption{Accuracy of the classical algorithms on the different datasets.}
\label{tab:results_classical}
\end{table}
  
As it can be seen from Table~\ref{tab:results_classical}, \tdtr is outperforming the other algorithms.
This is due to the fact that \squishc, \sttrace and \dr are designed to be less computationally expensive.

The performances of the \bwc algorithms on the AIS dataset can be found in Tables~\ref{tab:results_bwc_ais_10} and \ref{tab:results_bwc_ais_30}.
\begin{table}[ht]
\begin{tabular}{l r r r r r r}  
\toprule
window size (min)    &  120        &  60   &  15  &  5      & 0.5 \\      
points per window    &  800        &  400  &  100 &  33     & 4 \\        
\midrule                                                                  %
\bwcsquish               &  10.97   & 10.65 &  7.35 &  7.90  & 130.59\\    
\bwcsttrace              &  17.23   & 12.49 &  6.25 &  5.09  &  81.54 \\   
\bwcsttraceopt           &   1.49   &  1.53 &  1.72 &  4.62  & 108.39\\    
\bwcdr                   &  65.46   & 69.55 & 50.60 & 48.90  &  34.81 \\   
\bottomrule
\end{tabular}
\caption{Accuracy of the different \bwc algorithms when simplifying until $10\%$ of the AIS dataset for different sizes of time windows.}
\label{tab:results_bwc_ais_10}
\end{table}

\begin{table}[ht]
\begin{tabular}{l r r r r r r}  
\toprule
window size (min)    &  120        &  60    &  15  &  5      & 0.5 \\      
points per window    &  240        &  1200  &  300 &  100     & 12 \\        
\midrule                                                                  %
\bwcsquish               &  1.82    &  1.67 &  1.51  &  1.32 & 21.57 \\
\bwcsttrace              &  8.87    &  4.42 &  2.12  &  2.34 &  7.13 \\ 
\bwcsttraceopt           &  0.55    &  0.55 &  0.56  &  0.57 & 14.55 \\
\bwcdr                   & 19.60    & 19.48 & 12.15  & 10.36 &  9.60 \\
\bottomrule
\end{tabular}
\caption{Accuracy of the different \bwc algorithms when simplifying until $30\%$ of the AIS dataset for different sizes of time windows.}
\label{tab:results_bwc_ais_30}
\end{table}

Furthermore, we can notive from Tables~\ref{tab:results_bwc_ais_10} and \ref{tab:results_bwc_ais_30} that for large enough windows (between 15 and 120 minutes), \bwcsttraceopt is outperforming the other \bwc and classical algorithms.
This is due to the fact that the priority of the points is evaluated using the sample and the original trajectory. 
It can also be noticed that for small time windows, the performances of \bwcsquishc, \bwcsttrace and \bwcsttraceopt deteriorate. 
The deterioration is even drastic for 30 seconds time windows when keeping $10\%$ of the points.
This is due to the fact that these three algorithms compute the priority of a point according to both the previous and the next point in the sample.
Therefore, for small time windows, there will generally be less than 2 points per trajectory in the sample, making the removal of a point arbitrary and therefore leading to inaccurate simplifications.
On the other hand the performances of \bwcdr are more constant and even improve for smaller time windows.
This is due to the fact that \bwcdr only makes use of the previous one (or two) points to compute the priority of the currently processed point.
Therefore, even with small time windows, it will be able to compute the priorities correctly using points kept during the previous time windows.

As expected, it can also be noted that the average error of the improved version of \bwcsttraceopt is indeed smaller than the one of \bwcsttracec.
Surprisingly however, even \bwcsttrace outperforms the classical \sttrace algorithm.
One hypothesis is that this is due to \sttrace both assessing the priority of points using current simplified trajectory only and simultaneously comparing different trajectories of different natures. 
Therefore, trajectories with different sampling frequencies could be compressed simultaneously.
Trajectories with lower frequencies might fill up the priority queue as the priority of a point which is far apart in time from its neighbors in the sample will intuitively be higher than the one of a point close to its neighbors.
Restarting with an empty priority queue at frequent time interval might help mitigate this phenomenon.
\squish on the other hand, does not seem to suffer from this drawback. 
This might be due to their heuristic which counterbalance this effect by adding the priorities of points deleted from the sample.

The performances of the \bwc algorithms on the Birds dataset can be found in Tables~\ref{tab:results_bwc_birds_10} and ~\ref{tab:results_bwc_birds_30}.
\begin{table}[ht]
\begin{tabular}{l r r r r r r}  
\toprule
window size (days)       &  31        &  7   &  1  &  $1/4$      & $1/24$ \\      
points per window        &  5580      & 1260 &  180 &  45     & 8 \\           
\midrule                                                                       %
\bwcsquish               &  777   &  939 &  884 & 1061  & 3615\\    
\bwcsttrace              & 2780   & 2651 & 1144 & 1277  & 3096 \\   
\bwcsttraceopt           &  273   &  382 &  497 &  749  & 3437\\    
\bwcdr                   & 1997   & 1752 & 1677 & 1421  & 1314 \\   
\bottomrule
\end{tabular}
\caption{Accuracy of the different \bwc algorithms when simplifying until $10\%$ of the Birds dataset for different sizes of time windows.}
\label{tab:results_bwc_birds_10}
\end{table}

\begin{table}[ht]
\begin{tabular}{l r r r r r r}  
\toprule
window size (days)       &  31    &  7   &  1   &  $1/4$      & $1/24$ \\      
points per window        & 16740  & 3780 &  540 &  135         & 22 \\           
\midrule                                                                       %
\bwcsquish               &   77 & 104    & 108 & 126 & 4882\\    
\bwcsttrace              & 1245 & 707    & 245 & 247 & 6828   \\   
\bwcsttraceopt           &   32 &  50    &  60 &  77 & 4706 \\    
\bwcdr                   &  570 & 605    & 623 & 465 & 554  \\   
\bottomrule
\end{tabular}
\caption{Accuracy of the different \bwc algorithms when simplifying until $30\%$ of the Birds dataset for different sizes of time windows.}
\label{tab:results_bwc_birds_30}
\end{table}

Similar observations can be seen in Tables~\ref{tab:results_bwc_birds_10} and ~\ref{tab:results_bwc_birds_30} for the Birds dataset as for the AIS dataset.
Surprisingly, it can be seen that increasing the bandwidth from 8 to 22 points for the 1 hour time window lead to worse results for \bwcsquishc, \bwcsttrace and \bwcsttraceoptc.
This confirms the arbitrary simplification performed by these algorithms if there are not enough points for each trip in each time window. 


\subsection{Points distribution}

\begin{figure}[ht]
        \begin{center}
                \includegraphics[width=0.4\textwidth]{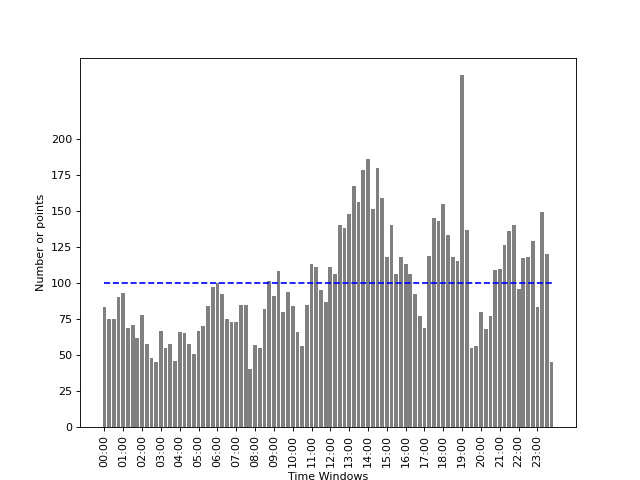}
        \end{center}
        \caption{Histogram of the quantity of points in different time-windows in samples obtained with \tdtrc.}
        \label{fig:hist_tdtr}
\end{figure}

\begin{figure}[ht]
        \begin{center}
                \includegraphics[width=0.4\textwidth]{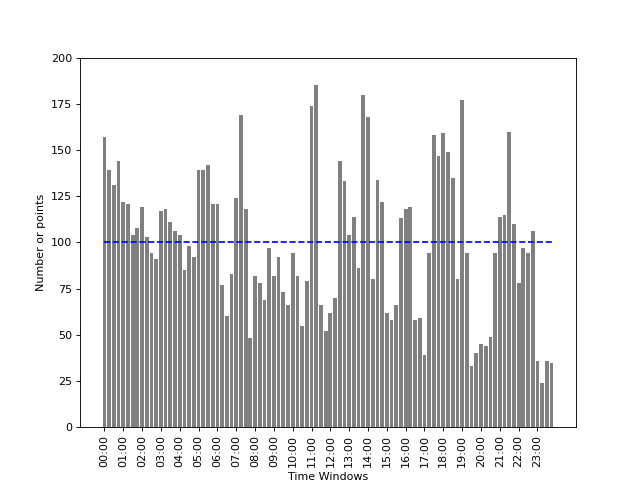}
        \end{center}
        \caption{Histogram of the quantity of points in different time-windows in samples obtained with \squishc.}
        \label{fig:hist_squish}
\end{figure}

\begin{figure}[ht]
        \begin{center}
                \includegraphics[width=0.4\textwidth]{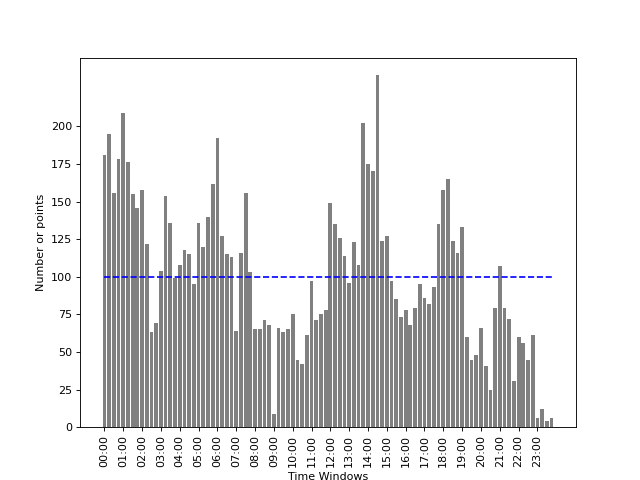}
        \end{center}
        \caption{Histogram of the quantity of points in different time-windows in samples obtained with \sttracec.}
        \label{fig:hist_sttrace}
\end{figure}

\begin{figure}[ht]
        \begin{center}
                \includegraphics[width=0.4\textwidth]{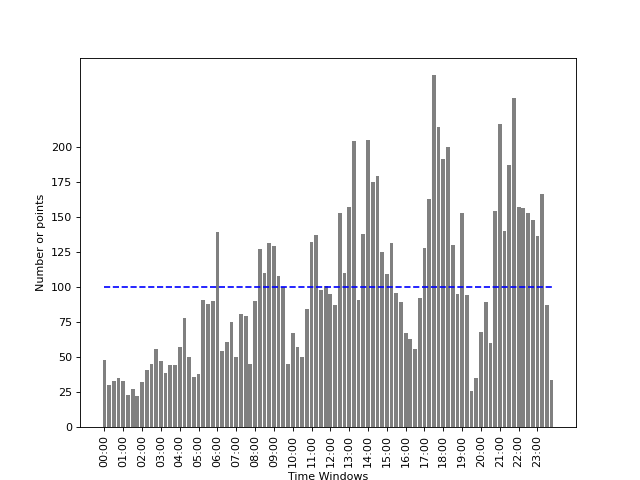}
        \end{center}
        \caption{Histogram of the quantity of points in different time-windows in samples obtained with \drc.}
        \label{fig:hist_dr}
\end{figure}

In this section, the time repartition of points conserved with classical compression algorithms will be illustrated.
This will be done by compressing the AIS dataset to $10\%$ of its original size and by analysing the time repartition of the points kept for each period of 15 minutes.
It will be shown that these algorithms do not produce an homogeneous time-partitioned results.
In this configuration, 100 points should be kept in each period in order to satisfy the bandwidth constrain.
The time repartition of simplified points for the \tdtrc, \squishc, \sttrace and \dr are illustrated in Figures \ref{fig:hist_tdtr}, \ref{fig:hist_squish},  \ref{fig:hist_sttrace} and \ref{fig:hist_dr}.
These figures consist in histograms representing the number of points remaining in all simplified trajectories during each period.

In each figure, the limit of 100 points is indicated with the blue dotted line.
These figures confirm the need of using different compression techniques in context with bandwidth constrains.

\section{Conclusion} \label{sec:ccl}

In this work, four variations of existing algorithms for the simplification of trajectories have been introduced.
These variations are aimed at being used in a situation with bandwidth limitations. 
The performances of the four algorithms have been studied for different sizes of time windows for two different datasets and for different compression rates.
While the more computationally intensive \bwcsttraceopt outperforms the other algorithms for the larger time windows, the performances of \bwcdr remain more stable with small time windows. 
Finally, the modified \bwcsquish provides satisfying results at a lower computational cost.

Several further improvements could still be considered.
First of all, this work extends three well known algorithms to a time windowed context. 
Different algorithms might also be considered for such an extension.
Furthermore, the presented algorithms could be further optimized. 
For instance the transition between time windows for the \bwcsquish as well as \bwcsttrace and \bwcsttraceopt could be improved.
Indeed, actually, all the last points of a trajectory in a window are assigned an infinity priority as there is no information accessible within the window with respect to the next points.
This is probably the main reason why  \bwcsquish, \bwcsttrace and \bwcsttraceopt perform so poorly when the number of points kept in a time window is low compared to the number of trips.
The priority of these last points could therefore be computed during the next time window, leading hopefully to more accurate results.
The \dr algorithm could also be modified in a different manner to satisfy bandwidth constrains instead of using a time-windowed approach with a priority queue.
For instance, the distance threshold could be modified in real time by the algorithm according to the number of points in the sample at a given time.

\begin{acks}
The research leading to the results presented in this paper
has received funding from the European Union’s funded
Project MobiSpaces under grant agreement no 101070279.

\end{acks}

\bibliographystyle{ACM-Reference-Format}
\bibliography{bibli.bib}


\begin{thebibliography}{16}


\ifx \showCODEN    \undefined \def \showCODEN     #1{\unskip}     \fi
\ifx \showDOI      \undefined \def \showDOI       #1{#1}\fi
\ifx \showISBNx    \undefined \def \showISBNx     #1{\unskip}     \fi
\ifx \showISBNxiii \undefined \def \showISBNxiii  #1{\unskip}     \fi
\ifx \showISSN     \undefined \def \showISSN      #1{\unskip}     \fi
\ifx \showLCCN     \undefined \def \showLCCN      #1{\unskip}     \fi
\ifx \shownote     \undefined \def \shownote      #1{#1}          \fi
\ifx \showarticletitle \undefined \def \showarticletitle #1{#1}   \fi
\ifx \showURL      \undefined \def \showURL       {\relax}        \fi
\providecommand\bibfield[2]{#2}
\providecommand\bibinfo[2]{#2}
\providecommand\natexlab[1]{#1}
\providecommand\showeprint[2][]{arXiv:#2}

\bibitem[\protect\citeauthoryear{Amigo, S{\'a}nchez~Pedroche, Garc{\'\i}a, and
  Molina}{Amigo et~al\mbox{.}}{2021}]%
        {amigo2021review}
\bibfield{author}{\bibinfo{person}{Daniel Amigo}, \bibinfo{person}{David
  S{\'a}nchez~Pedroche}, \bibinfo{person}{Jes{\'u}s Garc{\'\i}a}, {and}
  \bibinfo{person}{Jos{\'e}~Manuel Molina}.} \bibinfo{year}{2021}\natexlab{}.
\newblock \showarticletitle{Review and classification of trajectory
  summarisation algorithms: From compression to segmentation}.
\newblock \bibinfo{journal}{\emph{International Journal of Distributed Sensor
  Networks}} \bibinfo{volume}{17}, \bibinfo{number}{10} (\bibinfo{year}{2021}),
  \bibinfo{pages}{15501477211050729}.
\newblock


\bibitem[\protect\citeauthoryear{Denmark}{Denmark}{2021}]%
        {AISDenmark}
\bibfield{author}{\bibinfo{person}{AIS Denmark}.}
  \bibinfo{year}{2021}\natexlab{}.
\newblock \bibinfo{booktitle}{\emph{AIS Data}}.
\newblock Danish Maritime Authority.
\newblock
\urldef\tempurl%
\url{https://web.ais.dk/aisdata/}
\showURL{%
\tempurl}
\newblock
\shownote{Accessed on 27 December 2023.}


\bibitem[\protect\citeauthoryear{Dillen, Buroni, Le~Borgne, Determe, and
  Bontempi}{Dillen et~al\mbox{.}}{2020}]%
        {dillen2020mobi}
\bibfield{author}{\bibinfo{person}{Arnau Dillen}, \bibinfo{person}{Giovanni
  Buroni}, \bibinfo{person}{Yann-A{\"e}l Le~Borgne}, \bibinfo{person}{Karl
  Determe}, {and} \bibinfo{person}{Gianluca Bontempi}.}
  \bibinfo{year}{2020}\natexlab{}.
\newblock \showarticletitle{MOBI-AID: A Big Data Platform for Real-Time
  Analysis of On Board Unit Data.}. In \bibinfo{booktitle}{\emph{EDBT/ICDT
  Workshops}}.
\newblock


\bibitem[\protect\citeauthoryear{Douglas and Peucker}{Douglas and
  Peucker}{1973}]%
        {dp}
\bibfield{author}{\bibinfo{person}{David~H Douglas} {and}
  \bibinfo{person}{Thomas~K Peucker}.} \bibinfo{year}{1973}\natexlab{}.
\newblock \showarticletitle{Algorithms for the reduction of the number of
  points required to represent a digitized line or its caricature}.
\newblock \bibinfo{journal}{\emph{Cartographica: the international journal for
  geographic information and geovisualization}} \bibinfo{volume}{10},
  \bibinfo{number}{2} (\bibinfo{year}{1973}), \bibinfo{pages}{112--122}.
\newblock


\bibitem[\protect\citeauthoryear{Doulkeridis, Santipantakis, Koutroumanis,
  Makridis, Koukos, Theodoropoulos, Theodoridis, Kyriazis, Kranas, Burgos,
  Jimenez-Peris, Duarte, Sakr, Graser, Heistracher, Torp, Chrysakis,
  Orphanoudakis, Kapassa, Touloupou, Neises, Petrou, Karagiorgou, Catelli,
  Messina, {Corrales Compagnucci}, and Falsetta}{Doulkeridis
  et~al\mbox{.}}{2023}]%
        {mobispaces}
\bibfield{author}{\bibinfo{person}{Christos Doulkeridis},
  \bibinfo{person}{Georgios Santipantakis}, \bibinfo{person}{Nikolaos
  Koutroumanis}, \bibinfo{person}{George Makridis}, \bibinfo{person}{Vasilis
  Koukos}, \bibinfo{person}{{George S.} Theodoropoulos},
  \bibinfo{person}{Yannis Theodoridis}, \bibinfo{person}{Dimosthenis Kyriazis},
  \bibinfo{person}{Pavlos Kranas}, \bibinfo{person}{Diego Burgos},
  \bibinfo{person}{Ricardo Jimenez-Peris}, \bibinfo{person}{Mariana Duarte},
  \bibinfo{person}{Mahmoud Sakr}, \bibinfo{person}{Anita Graser},
  \bibinfo{person}{Clemens Heistracher}, \bibinfo{person}{Kristian Torp},
  \bibinfo{person}{{Ioannis Chrysakis} Chrysakis}, \bibinfo{person}{Theofanis
  Orphanoudakis}, \bibinfo{person}{Evgenia Kapassa}, \bibinfo{person}{Marios
  Touloupou}, \bibinfo{person}{Juergen Neises}, \bibinfo{person}{Petros
  Petrou}, \bibinfo{person}{Sophia Karagiorgou}, \bibinfo{person}{Rosario
  Catelli}, \bibinfo{person}{Domenico Messina}, \bibinfo{person}{Marcelo
  {Corrales Compagnucci}}, {and} \bibinfo{person}{Matteo Falsetta}.}
  \bibinfo{year}{2023}\natexlab{}.
\newblock \showarticletitle{MobiSpaces: An Architecture for Energy-Efficient
  Data Spaces for Mobility Data}.
\newblock \bibinfo{journal}{\emph{IEEE Big Data Service 2023}}.
\newblock


\bibitem[\protect\citeauthoryear{Juang, Oki, Wang, Martonosi, Peh, and
  Rubenstein}{Juang et~al\mbox{.}}{2002}]%
        {juang2002energy}
\bibfield{author}{\bibinfo{person}{Philo Juang}, \bibinfo{person}{Hidekazu
  Oki}, \bibinfo{person}{Yong Wang}, \bibinfo{person}{Margaret Martonosi},
  \bibinfo{person}{Li~Shiuan Peh}, {and} \bibinfo{person}{Daniel Rubenstein}.}
  \bibinfo{year}{2002}\natexlab{}.
\newblock \showarticletitle{Energy-efficient computing for wildlife tracking:
  Design tradeoffs and early experiences with ZebraNet}. In
  \bibinfo{booktitle}{\emph{Proceedings of the 10th international conference on
  Architectural support for programming languages and operating systems}}.
  \bibinfo{pages}{96--107}.
\newblock


\bibitem[\protect\citeauthoryear{Makris, Kontopoulos, Alimisis, and
  Tserpes}{Makris et~al\mbox{.}}{2021}]%
        {Markis2021}
\bibfield{author}{\bibinfo{person}{Antonios Makris}, \bibinfo{person}{Ioannis
  Kontopoulos}, \bibinfo{person}{Panagiotis Alimisis}, {and}
  \bibinfo{person}{Konstantinos Tserpes}.} \bibinfo{year}{2021}\natexlab{}.
\newblock \showarticletitle{A Comparison of Trajectory Compression Algorithms
  Over AIS Data}.
\newblock \bibinfo{journal}{\emph{IEEE Access}}  \bibinfo{volume}{9}
  (\bibinfo{year}{2021}), \bibinfo{pages}{92516--92530}.
\newblock
\urldef\tempurl%
\url{https://doi.org/10.1109/ACCESS.2021.3092948}
\showDOI{\tempurl}


\bibitem[\protect\citeauthoryear{McGillivary, Schwehr, and Fall}{McGillivary
  et~al\mbox{.}}{2009}]%
        {mcgillivary2009enhancing}
\bibfield{author}{\bibinfo{person}{Philip~A McGillivary},
  \bibinfo{person}{Kurt~D Schwehr}, {and} \bibinfo{person}{Kevin Fall}.}
  \bibinfo{year}{2009}\natexlab{}.
\newblock \showarticletitle{Enhancing AIS to improve whale-ship collision
  avoidance and maritime security}. In \bibinfo{booktitle}{\emph{OCEANS 2009}}.
  IEEE, \bibinfo{pages}{1--8}.
\newblock


\bibitem[\protect\citeauthoryear{Meratnia and de~By}{Meratnia and
  de~By}{2004}]%
        {meratnia2004spatiotemporal}
\bibfield{author}{\bibinfo{person}{Nirvana Meratnia} {and}
  \bibinfo{person}{Rolf~A de By}.} \bibinfo{year}{2004}\natexlab{}.
\newblock \showarticletitle{Spatiotemporal compression techniques for moving
  point objects}. In \bibinfo{booktitle}{\emph{Advances in Database
  Technology-EDBT 2004: 9th International Conference on Extending Database
  Technology, Heraklion, Crete, Greece, March 14-18, 2004 9}}.
  \bibinfo{publisher}{Springer}, \bibinfo{pages}{765--782}.
\newblock


\bibitem[\protect\citeauthoryear{Muckell, Hwang, Patil, Lawson, Ping, and
  Ravi}{Muckell et~al\mbox{.}}{2011}]%
        {squish}
\bibfield{author}{\bibinfo{person}{Jonathan Muckell},
  \bibinfo{person}{Jeong-Hyon Hwang}, \bibinfo{person}{Vikram Patil},
  \bibinfo{person}{Catherine~T Lawson}, \bibinfo{person}{Fan Ping}, {and}
  \bibinfo{person}{SS Ravi}.} \bibinfo{year}{2011}\natexlab{}.
\newblock \showarticletitle{SQUISH: an online approach for GPS trajectory
  compression}. In \bibinfo{booktitle}{\emph{Proceedings of the 2nd
  international conference on computing for geospatial research \&
  applications}}. \bibinfo{pages}{1--8}.
\newblock


\bibitem[\protect\citeauthoryear{Muckell, Olsen, Hwang, Lawson, and
  Ravi}{Muckell et~al\mbox{.}}{2014}]%
        {squishe}
\bibfield{author}{\bibinfo{person}{Jonathan Muckell}, \bibinfo{person}{Paul~W
  Olsen}, \bibinfo{person}{Jeong-Hyon Hwang}, \bibinfo{person}{Catherine~T
  Lawson}, {and} \bibinfo{person}{SS Ravi}.} \bibinfo{year}{2014}\natexlab{}.
\newblock \showarticletitle{Compression of trajectory data: a comprehensive
  evaluation and new approach}.
\newblock \bibinfo{journal}{\emph{GeoInformatica}}  \bibinfo{volume}{18}
  (\bibinfo{year}{2014}), \bibinfo{pages}{435--460}.
\newblock


\bibitem[\protect\citeauthoryear{Potamias, Patroumpas, and Sellis}{Potamias
  et~al\mbox{.}}{2006}]%
        {potamias2006sampling}
\bibfield{author}{\bibinfo{person}{Michalis Potamias}, \bibinfo{person}{Kostas
  Patroumpas}, {and} \bibinfo{person}{Timos Sellis}.}
  \bibinfo{year}{2006}\natexlab{}.
\newblock \showarticletitle{Sampling trajectory streams with spatiotemporal
  criteria}. In \bibinfo{booktitle}{\emph{18th International Conference on
  Scientific and Statistical Database Management (SSDBM'06)}}.
  \bibinfo{publisher}{IEEE}, \bibinfo{pages}{275--284}.
\newblock


\bibitem[\protect\citeauthoryear{Series}{Series}{2014}]%
        {series2014technical}
\bibfield{author}{\bibinfo{person}{M Series}.} \bibinfo{year}{2014}\natexlab{}.
\newblock \showarticletitle{Technical characteristics for an automatic
  identification system using time-division multiple access in the VHF maritime
  mobile band}.
\newblock \bibinfo{journal}{\emph{Recommendation ITU: Geneva, Switzerland}}
  (\bibinfo{year}{2014}), \bibinfo{pages}{1371--1375}.
\newblock


\bibitem[\protect\citeauthoryear{Shekhar, Gunturi, Evans, and Yang}{Shekhar
  et~al\mbox{.}}{2012}]%
        {shekhar2012spatial}
\bibfield{author}{\bibinfo{person}{Shashi Shekhar}, \bibinfo{person}{Viswanath
  Gunturi}, \bibinfo{person}{Michael~R Evans}, {and} \bibinfo{person}{KwangSoo
  Yang}.} \bibinfo{year}{2012}\natexlab{}.
\newblock \showarticletitle{Spatial big-data challenges intersecting mobility
  and cloud computing}. In \bibinfo{booktitle}{\emph{Proceedings of the
  Eleventh ACM International Workshop on Data Engineering for Wireless and
  Mobile Access}}. \bibinfo{pages}{1--6}.
\newblock


\bibitem[\protect\citeauthoryear{Stienen, Müller, Lens, Milotic, and
  Desmet}{Stienen et~al\mbox{.}}{2020}]%
        {birds}
\bibfield{author}{\bibinfo{person}{Eric W.~M. Stienen}, \bibinfo{person}{Wendt
  Müller}, \bibinfo{person}{Luc Lens}, \bibinfo{person}{Tanja Milotic}, {and}
  \bibinfo{person}{Peter Desmet}.} \bibinfo{year}{2020}\natexlab{}.
\newblock \bibinfo{booktitle}{\emph{LBBG\_JUVENILE - Juvenile lesser
  black-backed gulls ({\textit{Larus fuscus}}, Laridae) hatched in Zeebrugge
  (Belgium)}}.
\newblock
\urldef\tempurl%
\url{https://doi.org/10.5281/zenodo.5075868}
\showDOI{\tempurl}


\bibitem[\protect\citeauthoryear{Trajcevski, Cao, Scheuermanny, Wolfsonz, and
  Vaccaro}{Trajcevski et~al\mbox{.}}{2006}]%
        {trajcevski2006line}
\bibfield{author}{\bibinfo{person}{Goce Trajcevski}, \bibinfo{person}{Hu Cao},
  \bibinfo{person}{Peter Scheuermanny}, \bibinfo{person}{Ouri Wolfsonz}, {and}
  \bibinfo{person}{Dennis Vaccaro}.} \bibinfo{year}{2006}\natexlab{}.
\newblock \showarticletitle{On-line data reduction and the quality of history
  in moving objects databases}. In \bibinfo{booktitle}{\emph{Proceedings of the
  5th ACM international workshop on Data engineering for wireless and mobile
  access}}. \bibinfo{pages}{19--26}.
\newblock


\end{thebibliography}

%
% ...

\end{document}